\documentclass[a4paper,10pt]{edpsci} 
\usepackage{amsmath}
\usepackage{rotating}
\usepackage{graphicx}
\usepackage{epstopdf}
\usepackage[numbers]{natbib}
\usepackage{hyperref}
\usepackage{url}
\usepackage{amssymb}
\usepackage{lipsum}
\hypersetup{colorlinks=true,citecolor=blue,urlcolor=blue,linkcolor=blue}
\usepackage[ruled]{algorithm2e}
\begin{document}

\title{Impact of $\rm MAPbI_3$ Phase Transitions on Solar Cell Performance}

\titlerunning{$\rm MAPbI_3$ Phase Transitions Impacts on Cell Efficiency}

\subtitle{Everything you need to know about {\em ab-initio} methods in device performance}


\author{Ph. Baranek\inst{1,2}\correspondingauthor{\email{philippe.baranek@edf.fr}} 
\and
J.P. Connolly\inst{3}
\and
A. Gissler\inst{1,2,4} \and Ph. Schulz\inst{4} \and M. R\'erat\inst{5} \and R. Dovesi\inst{6}}


\authorrunning{Ph. Baranek {\em et al.}}


\institute{EDF R\&D, EDF Lab Paris-Saclay, Department SYSTEME, 7 boulevard Gaspard Monge, F-91120 Palaiseau, France \and
IPVF, Institut Photovoltaique d'Ile-de-France, 18 boulevard Thomas Gobert, F-91120 Palaiseau, France \and
Universit\'e Paris-Saclay, CentraleSupelec, CNRS, Laboratoire de G\'enie Electrique et Electronique de Paris, 11 rue Joliot Curie, F-91192 Gif-sur-Yvette, France \and 
Institut Photovoltaique d'Ile-de-France (IPVF), UMR 9006, CNRS, Ecole Polytechnique, IP Paris, Chimie Paristech, PSL, 18 boulevard Thomas Gobert, F-91120 Palaiseau, France \and 
Universit\'e de Pau et des Pays de l’Adour, E2S UPPA, CNRS, IPREM, 2 avenue du Pr\'esident Pierre Angot, F-64053 Pau, France \and
Accademia delle Science di Torino, via Accademia delle Science 6, I-10123 Torino, Italy}

\abstract{This paper presents a first step toward a pragmatic phenomenological multiscale approach to evaluate perovskite solar cell performance which determines material properties at the atomistic scale with first-principles calculations, and applies them in macro-scale device models. This work focuses on the $\rm MAPbI_3$ (MA = $\rm CH_3NH_3$) perovskite and how its phase transitions impact on its optical, electronic, and structural properties which are investigated at the first-principles level. The obtained data are coupled to a numerical drift-diffusion device model enabling evaluation of the performance of corresponding single junction devices. The first-principles simulation applies a hybrid exchange-correlation functional adapted to the studied family of compounds. Validation by available experimental data is presented from materials properties to device performance, justifying the use of the approach for predictive evaluation of existing and novel perovskites. The coupling between atomistic and device models is described in terms of a framework for exchange of optical and electronic parameters between the two scales. The obtained results are
systematically discussed in terms of first-principles levels of approximation performances.}
\maketitle
\section{Introduction}
\noindent Perovskite solar cell efficiency under standard test conditions has progressed extremely rapidly from 3.8 \% in 2009 to 27.3 \% in September 2024. Tandem efficiencies have furthermore breached the single-junction Shockley-Queisser efficiency limit, reaching 34.6 \% in June 2024 (LONGI, certified) \cite{Green_2025,NREL}. While this rapid efficiency increase is unmatched by any other technology, perovskite absorber materials still suffer stability issues linked to temperature, to volatile organic cations for the organic case and its reactivity to the air moisture among other issues. These issues are obstacles for industrial and societal applications.

Both air moisture and temperature induce phase transitions which degrade the performance and durability of perovskite solar cells (PSCs). The moisture leads to the appearance of a non-perovskite phase (the so-called $\delta$-black phase) which is optically inactive, while the temperature can lead to a rich sequence of phase transitions. Their impact concerns mainly the electronic properties, and the domains and surfaces stabilities of the different compounds.

In this work, we focus on the impact of $\rm MAPbI_3$ (MA = $\rm CH_3NH_3$) phase transitions on cell efficiency. For both organic and inorganic perovskites, these transitions are associated in particular to the existence of soft phonon modes which can locally generate phase instabilities \cite{walsh,Leguy2_2016}. They are linked to the lattice and halide octahedra deformations. However, for the organic case, another factor has to be taken into account which is the nonlocal ordering of the organic moieties inside the lattice through the different phase transitions which yields lattice distortions. These free rotations of the cations and induced distortions directly impact the band gap, and absorption and transport properties \cite{Even_2014,Gao_2016}. On the first-principles level, modeling these types of materials with a static approach remains a challenge since the MA contribution to the phase transition is linked to its dipole moment. Li and co-workers \cite{JLi_2016,JLi_2018} have shown that the inorganic-framework deformation depends on the orientation of the organic cation which directly influences the stability of the hybrid perovskites and requires a multiscale approach to obtain a good description of their properties; the MA dipole moment induces polarized dipole order in $\rm MAPbI_3$ at the nano-micro scales in order to minimize the electrostatic dipole-dipole energy or interaction \cite{Frost_2014,Frost2_2014,Jarvil_2018}.

From an experimental point of view, one of the main consequences of these various trends is the difficulty in obtaining an accurate characterization of its different phases due to the determination of the MA atomic positions inside the $\rm PbX_3$ lattice. Since the measurements are mainly performed with X-Ray diffraction, their positions are ill defined. Therefore, for the cubic phase the assumption  is to consider that MA lies in the center of the cubic cell. Moreover, with this description $\rm MAPbI_3$ is in a ferroelectric phase which will influences the characterization of its optoelectronic properties. However, this "local" representation stays the most commonly used in the field of the first-principles modeling of the organic-inorganic hybrid perovskites.

The evaluation of solar cell performance relies on the determination of the optical response of the absorber. Nowadays, a lot of theoretical studies used the determined optical response at different levels of approximations as input for device model without considering the problem of their reliability which tends to lead to unrealistic performances. Depending of the absorber's nature, their calculation can need the use of "sophisticated" methods such as the quasi-particle approaches, for instance. To the best of our knowledge, there are only few systematic attempts of determination of the $\rm MAPbI_3$ optical responses \cite{Leguy_2016,Demchenko_2016}. All of them are based on a "local" representation of the perovskite and illustrate the difficulty to obtain a dielectric response in, at least, qualitative agreement with experiment, even when performed with "state of the art" methods. However, Demchenko and co-workers \cite{Demchenko_2016} show that the use of a tuned screened hybrid HSE functional \cite{Krukau2006} in order to obtain a good reproduction of the band gap of the material leads to significant improvements of the dielectric response in the TDDFT framework.

 In this paper, using a theoretical multiscale approach, we illustrate how phase transitions can impact the performance of solar cells. This modeling couples atomistic scale first-principles calculations to device scale numerical models. The coupling between atomistic and device models is described in detail by presenting a framework for exchange of optical, and electronic parameters between the two scales.
 
This approach is based on a crystallographic description of $\rm MAPbI_3$ which allows to take the MA ordering into account. We first describe this crystallographic model. At the first-principles level, it is used to determine a hybrid exchange-correlation functional adapted to the $\rm MAPbX_3$ (x = Cl, Br and I) family of perovskites. The influence of the crystal phases on the electronic and dielectric properties of $\rm MAPbI_3$ is then systematically investigated. The corresponding band gaps, electron affinities and dielectric responses serve as input data to the device model which integrates these data in the absorber of a standard perovskite solar cell design \cite{Green_2014} which we will not detail here. This device model yields the corresponding solar cell performance allowing evaluation of the impact of materials configurations at the atomistic scale on device performance and stability. The obtained results are systematically commented in terms of first-principles approaches performances. 
\section{Methodological aspects}
\label{Methods}
\subsection{Crystallographic model}
\label{crystallo}
\noindent Before discussing the hybrid functional which best describes this family of materials, a
crystallographic structure allowing to obtain the ordering of the methalominium (MA = $\rm CH_3NH_3^+$) moities inside the lattice through the different phase transitions has to be defined.

\begin{figure}[h!]
\centering
\vspace{-1.3cm}
\includegraphics[scale=0.375]{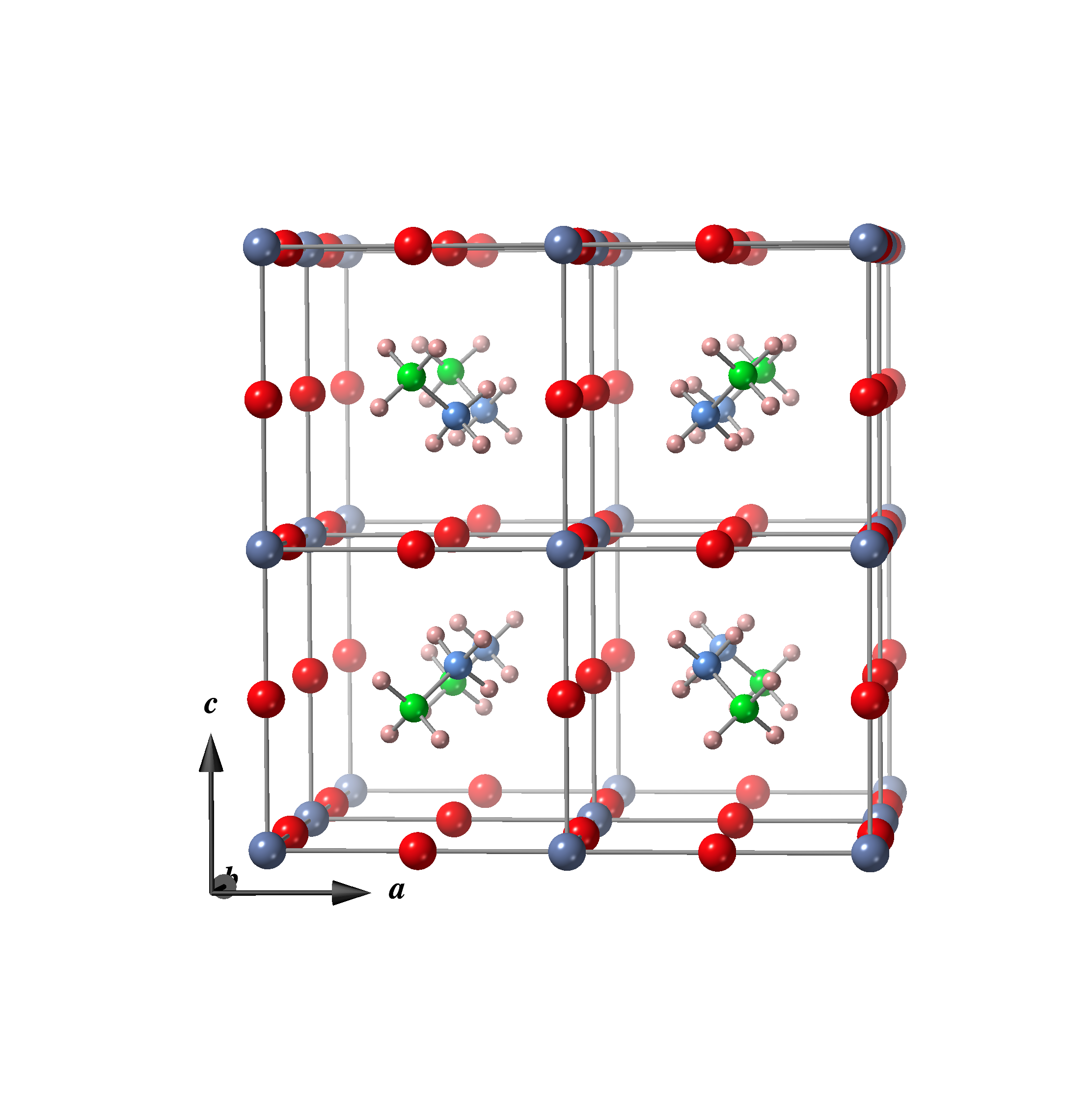}\\
\vspace{-1.2cm}
a) Cubic - $Pm\bar{3}m$\\
\vspace{-0.6cm}
\includegraphics[scale=0.1]{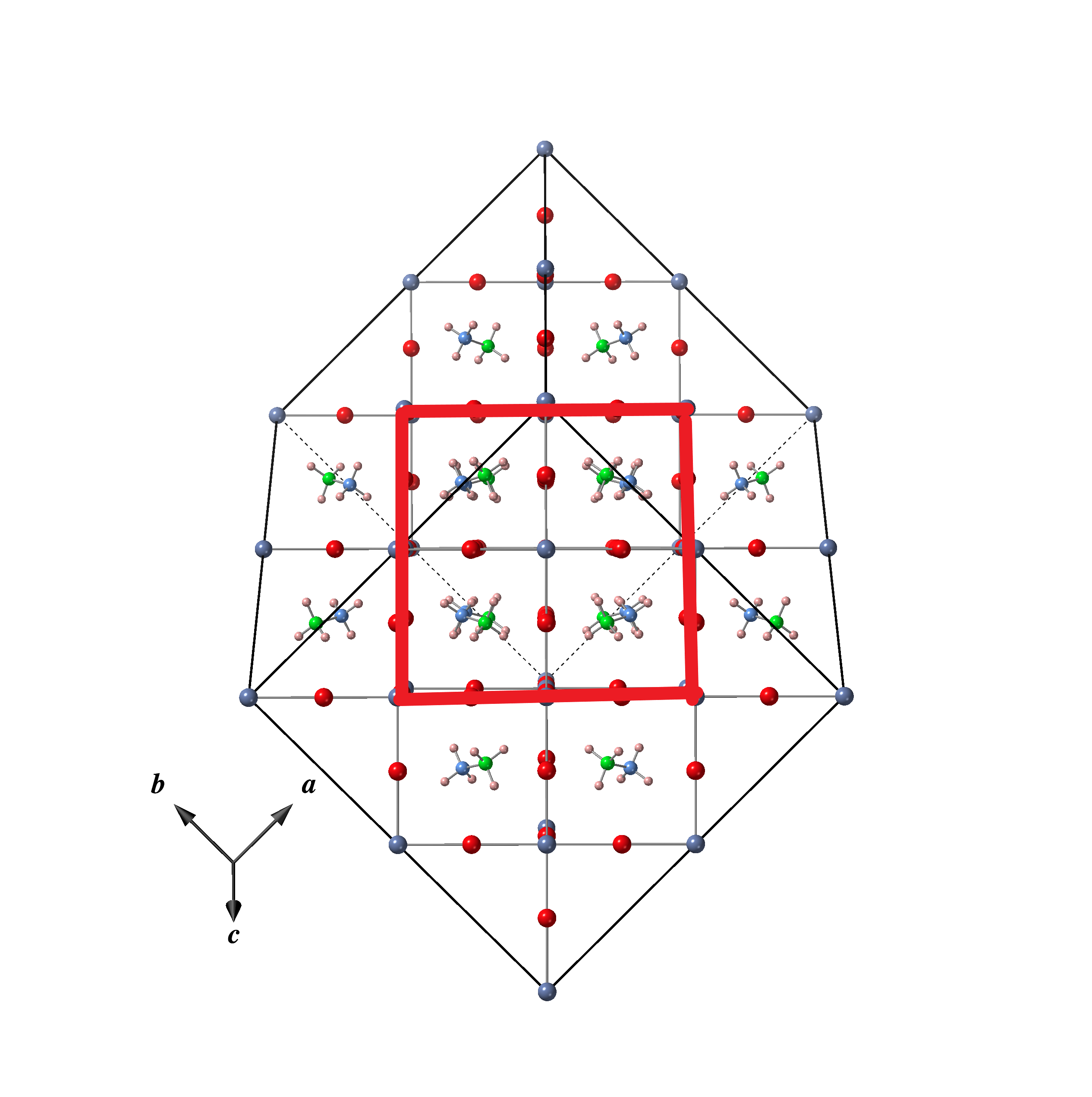}\\
\vspace{-0.6cm}
b) Tetrahedral - $I4/mcm$\\
(in red, the tetragonal $P4/mmm$ cell)\\
\vspace{-0.6cm}
\includegraphics[scale=0.35]{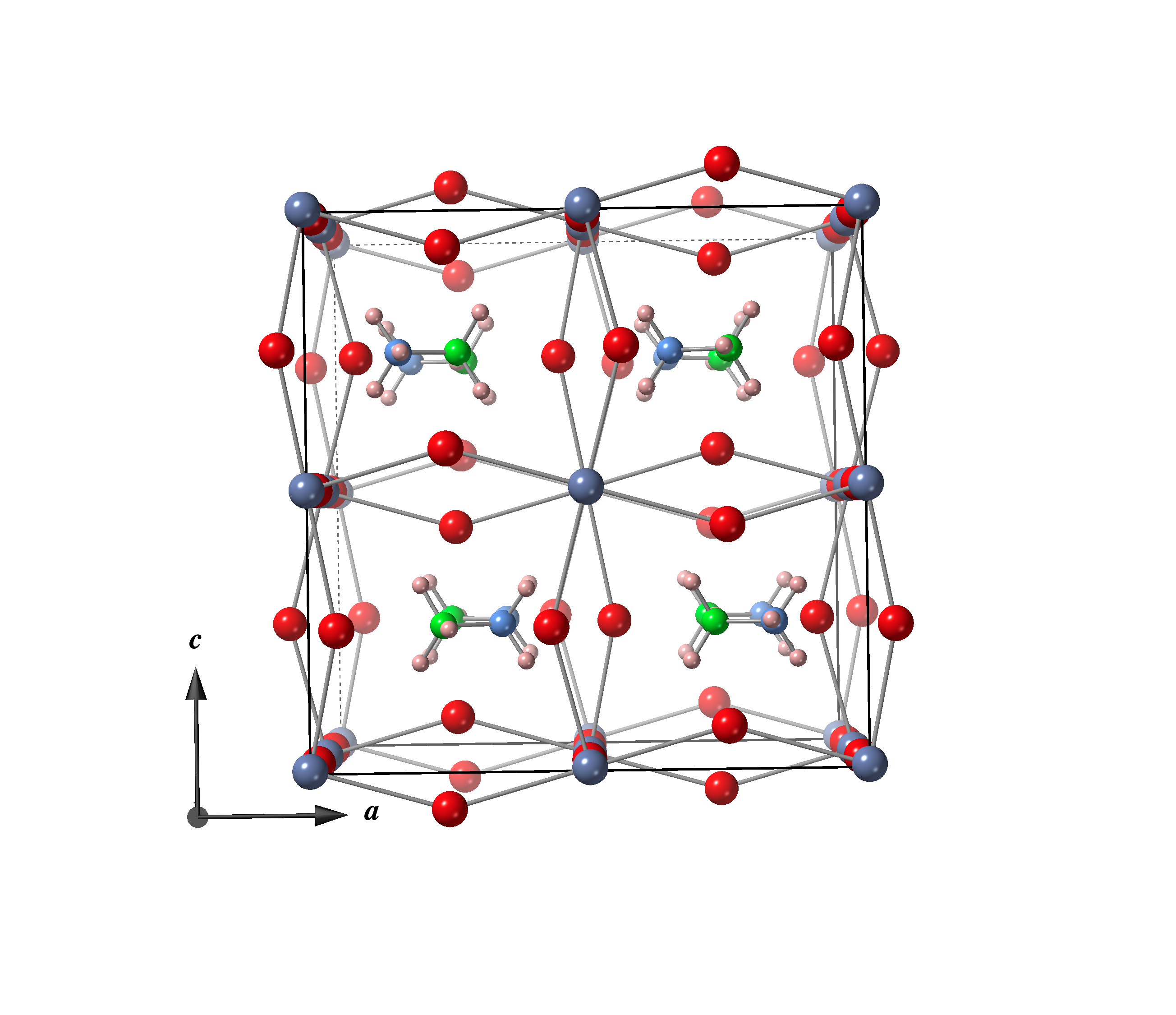}\\
\vspace{-0.9cm}
c) Orthorhombic - $P222_1$\\
\caption{Crystallographic structures of CH$_3$NH$_3$PbI$_3$ as used in this work. Spheres in gray -- Pb, red -- I, blue -- C, green -- N, and in pink -- H.\label{fig:crystal_structure}}
\end{figure}

 The $\rm MAPbX_3$ perovskites possess different phase transitions depending on temperature \cite{Poglitsch_1987,Schuk_2018,Dintakurti_2022}.These start with the high-temperature cubic perovskite structure ($Pm\bar{3}m$, stable above 330, 236 and 177K for $\rm MAPbI_3$, $\rm MAPbBr_3$ and  $\rm MAPbCl_3$, respectively) and goes to different orthorhombic ones ($Pnma$ or $Pna2_1$ below 161 and 149K for $\rm MAPbI_3$ and $\rm MAPbBr_3$, respectively, or $P222_1$ below 172K for $\rm MAPbCl_3$) via tetragonal phases (such as $I4/mcm$ for $\rm MAPbI_3$ or $P4/mmm$ for $\rm MAPbBr_3$ and $\rm MAPbBr_3$). These transitions from the cubic to tetragonal or orthorhombic phases are due to both factors: the tilting and distortion of the $\rm PbX_6$ octohedra, and, the ordering of the MA moities inside the lattice through the different phase transitions (as illustrated by figure \ref{fig:crystal_structure}).\\
Experimentally, the diversity of descriptions of a same phase, such as orthorhombic, comes from the difficulty in obtaining an accurate characterization of the MA atomic positions inside the $\rm PbX_3$ lattice: since the measurements are mainly performed with X-Ray or neutronic diffractions \cite{Poglitsch_1987,Dintakurti_2022,Lopez22,Chi_2005,Whitfield_2016,Caputo_2019,Maculan_2015,Li_2016,Zhu_2022,Mannino_2020,Lopez_2020,Page_2016,Lopez_20,Das_2022}, the positions of the MA moiety are ill or not defined. Therefore, for the $Pm\bar{3}m$ cubic phase the commonly used assumption is to consider MA as an intrinsic chemical entity which lies in the center of the cubic cell. However, this is not consistent from a crystallographic point of view: since the MA point group is $C_{3v}$, depending of the direction of the C--N bond inside the cubic (or {\em pseudo}-cubic) $\rm PbX_3$ lattice, the corresponding space groups are $C_{3v}$ ($R3m$), $C_2$ ($P2$) and $C_s$ ($Pm$) for the C--N bond oriented along the [1,1,1], [0,1,1] and [0,0,1] directions of the cubic cell, respectively. Moreover, MA possessing a strong dipole moment of 2.5 debye, with the commonly used description this material is obligatory in a ferroelectric phase, not observed experimentally, which might lead to a wrong theoretical characterization of its optoelectronic properties. Equivalent descriptions are used for the other phases \cite{Mosconi_2016,Mosconi_2014,Ponce_2019_2,Ponce_2019,Filip_2015,Filip_2014} but stay not consistent with the experimentally determined crystallographic space groups.

 The perovskite-type materials are known to develop structure instabilities, which can be grasped as softening of particular phonon modes (see, for instance, Ref. \cite{walsh,Leguy2_2016}). The most typical is the zone-boundary instability, which manifests itself as a tilting of the $\rm PbX_6$ octahedra leading to the cubic to orthorhombic phases. Typically, the cubic perovskite phase of these compounds undergoes distortions related to unstable phonon modes at the $\Gamma$, $X$, $M$ and $R$ points of the Brillouin zone.

 In order to obtain a structural description consistent with the crystallographic space groups of the different identified phases but also taking into account the zone-boundary instabilities, a supercell approach has to be considered. An equivalent supercell description of Li {\em et al.} \cite{JLi_2016,JLi_2018} which allows us to take the octahedra tilting  into account as well as the ordering of the MA moities has been adopted; a 2 $\times$ 2 $\times$ 2 supercell has been used for the $P$--primitive cubic, tetrahedral and orthorhombic phases, and a 2$\sqrt{3}$ $\times$ 2$\sqrt{3}$ $\times$ 2$\sqrt{3}$ supercell for the $I$--conventional cell of the tetrahedral phase. Figure \ref{fig:crystal_structure} gives the obtained atomic structures of the different phases (the full descriptions of the crystal's lattice are given in table SI).
\subsection{First-principles approach}
\label{abinitio}
\noindent First-principles calculations have been performed with the CRYSTAL code \cite{crystal17,crystal171,web_crys}. This program enables to solve both the Hartree–Fock (HF) and the Kohn–Sham (KS) systems of equations, combining them within a hybrid scheme.

 This work uses a hybrid exchange-correlation functional optimized to yield description of the structural, electronic, and dynamic properties of $\rm MAPbX_3$ (X = Cl, Br and I) in good agreement with experiment. This pragmatic 
approach allows us to define the Hamiltonian best suited for describing the 
properties of a given family of materials. It has recently been used to efficiently 
 study perovskite surface properties \cite{Mishra_2021}, the humidity-induced degradation products of halide perovskites \cite{Mejaouri_2024}, the structural and electronic properties of wide band gap hybrid perovskites \cite{Ory_2024} and the influence of alkali metals on the properties of chalcopyrites \cite{Lafond_2020}. In this work, the Hamiltonian (denoted as PBEx) combines 19\% of HF exact exchange with the PBE exchange correlation functional \cite{Perdew_2008}. Benchmark calculations, with corresponding values of equilibrium lattice parameters and band gaps, are reported in tables \ref{tab:hamiltonian}, \ref{tab:calc_crystal_MAPbI} and, SII: the obtained mean absolute average errors on the lattice parameters and band gaps of their most commonly characterized different phases are 2\% and 5\%, respectively, with respect to the available experimental data.

As regards the atomic basis sets, all-electron Gaussian-type-functions (GTFs) have been used for H, C, and N while the Stuttgart-Dresden fully relativistic pseudopotentials have been adopted for I  \cite{Peterson_2006} and Pb \cite{Metz2000}. The – 4$s$, 4$p$, 4$d$, 5$s$ and 5$p$ – electrons of I and the – 5$s$, 5$p$, 5$d$, 6$s$ and 6$p$ – electrons of Pb are treated as valence electrons. The basis sets of I and Pb are described in detail in \cite{Mishra_2021} and those of H, C and N in \cite{Ory_2024,web_crys_basis}.

For the evaluation of the Coulomb and exchange series within CRYSTAL,
the truncation thresholds for the bielectronic integrals, as defined in the CRYSTAL manual \cite{crystal171}, were set to $10^{-8}$, $10^{-8}$, $10^{-8}$, $10^{-8}$ and $10^{-16}$ Ha. The calculations have been performed with a
10 $\times$ 10 $\times$ 10 Monkhorst-Pack $k$-points meshes \cite{monkhorst}. The convergence criteria on total energies (and for the determination of frequencies )
were $10^{-9}$ ($10^{-12}$) Ha. The atomic displacements threshold in the atomic
relaxation was set to $1.8 \cdot 10^{-3}$ Bohr, and that for (converged) forces -- to $4.5 \cdot 10^{-4}$ Ha/Bohr. With these computational conditions, the obtained data can be considered fully converged.

The dielectric responses have been determined with a developer version of the CRYSTAL code. An 18 $\times$ 18 $\times$ 18 Monkhorst-Pack $k$-point mesh has been used to determine the real ($n$) and complex ($k$) refractive indices of the different phases. The affinities have been evaluated combining the calculation of the valence band edge energy based on the Janak theorem \cite{Lany_2009} and of the void energy via the determination of the macroscopic potential through the (001) surface of the cubic phase.

 The resulting band gaps, affinities and optical indices serve as input data to the device model which yields the performance of solar cells.
\subsection{Device model}
\label{drift}
%
%
%
%
%
%
\noindent The perovskite solar cell (PSC) device scale modelling  uses finite element numerical modelling via SILVACO’s ATLAS simulator \cite{silvaco}. This uses the well known drift diffusion model which consists of solving the current, continuity, and Poisson equations which we need not repeat here. The atomistic modelling being focussed on electronic and optical properties, and therefore not concerned with ion migration in this work, this is not included in the device model.

\begin{figure}[h!]
\centering
\includegraphics[width=245pt]{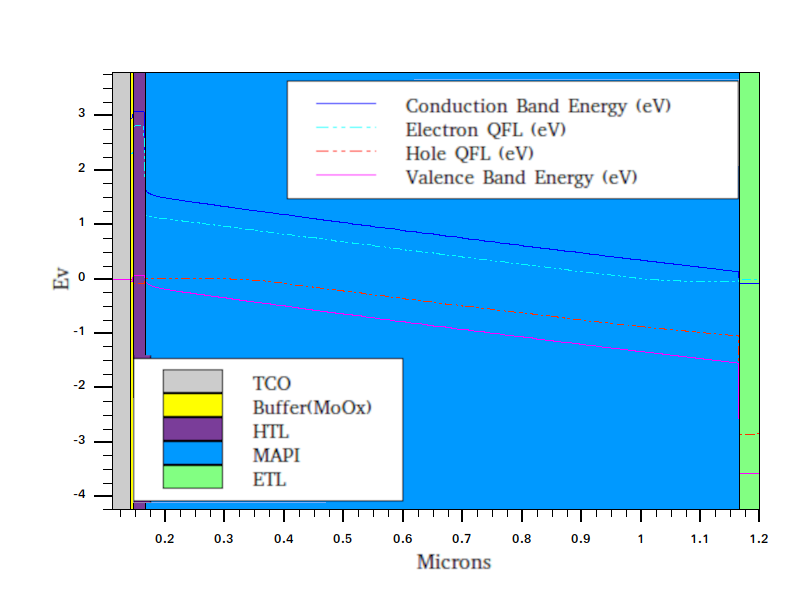}\\
\caption{Perosvkite cell layer structure and calculated band profile for experimental parameter input values to the device model.\label{fig:psc_structure}}
\end{figure}

 The device structure we choose is an ``inverted'' $\rm p-i-n$ polarity which is standard in the field (\cite{docampo_2013}, \cite{nazeeruddin_2015}) and is illustrated in figure \ref{fig:psc_structure}. This inverted design consists of electron transport layer (ETL), an intrinsically perovskite, where we make the common assumption of negligible doping, and hole transport layer (HTL) with contacts on a glass substrate. The front surface consisists of a transparent conducting oxide  and a thin MoOx buffer layer protecting the perovskite. The optical properties are simulated with a transfer matrix methodology and diffusive optics to simulate imperfectly planar surfaces of typical structures.  The model outputs include all the usual performance figures of merit such as the conversion efficiency as presented in the results section.\\ 
\begin{table}[h!]
\caption{\label{tab:device}Full set of device scale drift-diffusion (DD) model inputs from atomistic scale density functional scale (DFT) model. This study uses a subset which are band parameters and optical functions.}
\begin{center}
\begin{tabular}{cc}
\hline
Parameter&Definition\\
\hline
\multicolumn{2}{c}{Band parameters}\\
\hline
$E_C$, $E_V$& Conduction and valence band edges\\
$E_g$& Band gap\\
N$_C$, N$_V$&Conduction and valence band \\
$m_{e, h}^*$&Electron and hole \\
&effective masses\\
$\chi$&electronic affinity\\
&\\
\multicolumn{2}{c}{Optical data}\\
\hline
$\varepsilon_r$, $\varepsilon_i$&Real and imaginary permittivities\\ 
&related to complex refrative index\\
$n$, $k$&Real and imaginary\\
& refractive indices\\
&\\
\multicolumn{2}{c}{Diffusion paramters}\\
\hline
$\tau_{SRH}$	& Electron and hole charge neutral \\
& and depletion layers \\
& Shockley-Read-Hall lifetimes\\
$\mu$ & Carrier mobility,\\
& majority and minority, \\
& electron and hole\\
D$_N$, D$_P$&	Hole and electron \\
& diffusion coefficients\\
C$_A$&	Auger coefficient\\
\hline
\end{tabular}
\end{center}
\end{table}
\indent The multiscale coupling consists of identifying device level parameters which can be provided by atomistic scale density functional theory materials models. The full list of materials parameters required are summarised in table \ref{tab:device}, grouped in terms of band, optical, and transport parameters.\\
The first of these are the work functions of ETL and HTL transport layers. The difference between these work fonctions provides the build-in field across the absorber, analogous to the field provided by p-doped and n-doped charge neutral layers in traditional solar cells. The photocurrent, however, is provided only by collection of carriers photogenerated in the perovskite intrinsically doped absorber. This is because ETL and HTL layers being predominantly conductors rather than charge-neutral diffusive transport dominated layers found in traditional solar cells, they contribute essentially no photocurrent.

Concerning the perovskite absorber where photocurrent is generated, the slope of the conduction and valence bands in the perovskite absorber (figure  \ref{fig:psc_structure}) indicates this is a space charge region. Photogenerated  carrier transport in this region is therefore dominated by drift under the electric field, and is not sensitive to carrier diffusion parameters in the absorber. As a result, device performance is dominated by optical and band parameters of ETL, perovskite, and HTL for the purposes of comparing perovskite materials. We therefore make the assumption that carrier transport is independent of perovskite composition and use the same standard values for these in modelling of the different phases.

Furthermore, objective of the device scale modelling is to allow the evaluation of potential performance of different perosvkite phases, and evaluation of the accuracy of atomistic modelling output. In this study, therefore, we limit the multiscale interaction to optical and band structure parameters, and use standard values of all other parameters available in the literature which we have reported in previous work \cite{Connolly_2020}.
\section{Results and discussion}
\begin{table*}[h!]
\caption{\label{tab:hamiltonian}Lattice parameter $a$ and (direct) optical band gap $E_g$ for the cubic $\rm MAPbI_3$ perovskite (space group $Pm\bar{3}m$, see Fig. \ref{fig:crystal_structure}a)). Calculated results of the present work are obtained with the
specially tuned hybrid functional PBEx (see text), PBE exchange-correlation functional \cite{Perdew_2008} and the HSE screened hybrid functional \cite{Krukau2006}. A selection of theoretical and experimental data are given for comparison. $a$ and $E_g$ for obtained for the rhombohedral single cell are also provided.}
\begin{center}
\begin{tabular}{lccccccc}
\hline
&&\multicolumn{3}{c}{Present work}&&\multicolumn{2}{c}{Other works}\\
\cline{3-5}
\cline{7-8}
Property&&PBEx&PBE&HSE&&Theory$^a$&Expt.\\
\hline
$a$ (in \AA)&&6.368&6.383&6.358&&&6.329$^b$, 6.306$^c$\\
$E_g$ (in eV)&&1.68&0.92&1.36&&1.27$^d$, 1.67$^e$&1.60$^j$, 1.70$^k$\\
&&&&&&1.48$^f$&1.62 (1.50--1.69)$^l$\\
&&&&&&1.08 (0.94)$^g$, 1.33 (1.26)$^h$, 1.65 (1.64)$^i$&\\
\hline
\end{tabular}
\end{center}
\footnotesize
$^a$Results obtained at the experimental lattice constants of Ref. \cite{Poglitsch_1987,Baikie_2013}; $^b$Ref.\cite{Poglitsch_1987} XRD; $^c$Ref.\cite{Whitfield_2016} neutron powder diffraction on CD$_3$ND$_3$PbI$_3$; $^{d \text{ and } e}$Ref.\cite{Brivio_2014} determined with $G_0W_O$ and $QSGW$ ($QS$ for quasi-particle self-consistent) approaches including spin-coupling and using the LDA electronic properties, respectively; $^{f}$Ref.\cite{Ahmed_2014,Ahmed_2015} determined with a $G_0W_O$ approach including spin-coupling and using the PBE electronic properties; $^{g \text{, } h \text{ and } i}$Ref. \cite{Lepper_2019} calculated with a $GW_0$ ($G_0W_0$) including spin-orbit coupling approach using electronic properties determined at the PBE, HSE and PBE0 \cite{Adamo_1999} levels, respectively; $^j$Ref. \cite{Caputo_2019}; $^k$Ref. \cite{Li_2016}; $^l$average value of experimental data from Table 2 and between parentheses range of variation of the band gap with different materials formings and  measurement techniques cited in Table 1 of Ref. \cite{Das_2022}, respectively.
\end{table*}
\subsection{Crystals structures and band gaps evaluation}
\noindent The tables \ref{tab:hamiltonian} and \ref{tab:calc_crystal_MAPbI} provide the results obtained for the geometries and band gaps of different $\rm MAPbI_3$ phases. They are compared to available experimental data, and, determined band gaps at the state of the art many-body perturbation theory based on Green's function ($GW$) approximation using electronic properties calculated with different functionals at various DFT levels of approximations (a detailed review on the many-body approaches performances applied to perovskites is provided in Ref. \cite{Filip_2024}). The combination of the crystallographic description, presented in section \ref{crystallo}, and the hybrid functional defined in section \ref{abinitio} provides results on these properties consistent with experiment and with a more homogeneous qualitative and quantitative description than the most commonly used screened hybrid functional 
HSE \cite{Krukau2006} for semiconductors, or, the most sophisticated methods based on the $GW$  approximation: this can be judged by the comparison done in both tables with experiment: for $\rm MAPbI_3$, the obtained average errors on the lattice parameters and band gaps are 1.6 ($\pm$ 0.6) and 5. ($\pm$ 3) \%, respectively.

The table \ref{tab:hamiltonian} summarizes on the cubic phase example the obtained general trends:
\begin{itemize}
    \item[i)] The performance of the different functionals which we obtain with our crystallographic description of $\rm MAPbI_3$ are coherent with the ones well established in the literature: The lattice parameters and band gaps determined at the PBE level are overestimated and underestimated with respect to experiment, respectively; the use of hybrid functionals, such as PBEx or HSE, allows the correction of PBE discrepancies. The best agreement is obtained with our PBEx functional.
    \item[ii)] The band gaps determined at the many-body level are sensitive to further factors:\\
    - The level of sophistication of the $GW$ approach: The best correction is reached with the approach based on quasi-particle self-consistent methods.\\
    - The choice of the functional level to create the basic electronic properties on which are applied the $GW$ approximations. The best agreement with experiment is obtained for $GW$ band gaps determined from hybrid functionals, the $GW$ methods allowing the correction of LDA and GGA discrepancies but without reaching the same level of accuracy.\\
    - The choice of the phase description: The many-body approaches do not permit the use of structural data optimized at the same level of approximation. $GW$ calculations can be made using either experimental structures or structures optimized with other functionals. However, for hybrid perovskites, the amplitude of the lattice deformations depends on the orientation of the MA$^+$ cation \cite{JLi_2016}, which has a significant impact on the obtained results, as illustrated in Table SIII: this shows that imposing a cubic lattice with undistorted PbX$_6$ octahedra and with the MA moity oriented in a given direction can lead to variations of the band gaps of more than 50 \% with respect to the fully optimized cells.
\end{itemize}

The table \ref{tab:calc_crystal_MAPbI} presents the evolution of the relative stability ($\Delta H$) of different identified tetragonal and orthorhombic phases with respect to the cubic perovskite, and, their band gaps ($E_g$). To understand the different trends described below, we remind that:
\begin{itemize}
\item[i)] The halides perovskites have a specificity compared to other semiconductors or oxides perovskites. They possess an ”inverted” band structure (in the words of Huang and Lambrecht \cite{Huang_2013}): only the lead and iodine $s$ and $p$ states contribute to the top of the valence band and the bottom of the conduction bands; Pb cations have an occupied $6s$ orbital (lone pair) interacting with the in-plane I-$5p$ orbitals to form the top of the valence band when the Pb-$6p$ and I-$5p$ states form the bottom of the conduction bands \cite{Olthof_2016}. Then the band gaps are directly impacted by the halide octahedra distortions which increase the dispersion of the top of valence bands and of the bottom of the conduction bands, opening the band gaps and increasing the anysotropy of the perovskites especially their optoelectronic properties \cite{Chen_2025} (as illustrated by figure SI).
\item[ii)] The MA contribution to the phase transition is linked to its dipole moment which induces polarized dipole order in $\rm MAPbI_3$ at the nano-micro scales in order to minimize the electrostatic dipole-dipole energy or interaction \cite{Frost_2014,Frost2_2014,Jarvil_2018}.
\end{itemize}
\begin{table*}[!h]
\begin{center}
\caption{Calculated lattice parameters ($a$, $b$ and $c$ in {\AA}), band gap ($E_g$ in eV) and energy difference with the cubic perovskite ($\Delta H$ in meV/MAPbX$_3$) for the different phases of $\rm MAPbI_3$. Except for the tetragonal $I4/mcm$ phase, the geometrical data are expressed in the pseudo-cubic referential. A selection of available theoretical and experimental data are given for comparison.\label{tab:calc_crystal_MAPbI}}
\begin{tabular}{lccccc}
\hline
Phase&&This work&\multicolumn{2}{c}{Other works}\\
\cline{4-5}
&&&Theor.$^a$&Exp.\\
\hline
Cubic&$a$&6.368&&6.329$^b$, 6.306$^c$\\
$Pm\bar{3}m$&$E_g$&1.68&1.08$^d$, 1.65$^e$&1.60$^l$, 1.70$^m$\\
&&&1.67$^f$&1.62 (1.50--1.69)$^n$\\
\hline
Tetragonal&$a$&9.013&&8.855$^b$, 8.799$^o$, 8.857$^p$\\
$I4/mcm$&&({\em 9.029})&&\\
&$c$&12.629&&12.659$^b$, 12.688$^o$, 12.651$^p$\\
&&({\em 12.650})&&\\
&$E_g$&2.17&1.64$^g$, 1.67$^h$&1.60$^p$\\
&&({\em 1.81})&&1.54 (1.46--1.62)$^n$\\
&$\Delta H$&-200.&&\\
&&(-135.)&&\\
\hline
Orthorhombic&$a$&6.223&&6.053$^o$, 6.069$^q$ \\
$Pnma$&$b$&6.241&&6.231$^o$, 6.254$^q$ \\
&$c$&6.477&&6.294$^o$, 6.296$^q$\\
&$E_g$&2.35&1.57$^i$, 1.71$^j$&\\
&&&1.79$^k$&\\
&$\Delta H$&-267.&&\\
\hline
Orthorhombic&$a$&6.202&&6.067$^b$\\
$Pna2_1$&$b$&6.226&&6.266$^b$\\
&$c$&6.487&&6.310$^b$\\
&$E_g$&2.24&&\\
&$\Delta H$&-298.&&\\
\hline
Orthorhombic&$a$&6.190&&\\
$P222_1$&$b$&6.230&&\\
&$c$&6.492&&\\
&$E_g$&2.25&&\\
&$\Delta H$&-310.&&\\
\hline
\end{tabular}
\end{center}
\footnotesize
$^a$Results obtained at the experimental lattice constants of Ref. \cite{Poglitsch_1987,Schuk_2018,Dintakurti_2022,Baikie_2013}; $^b$Ref.\cite{Poglitsch_1987} XRD; $^c$Ref.\cite{Whitfield_2016} neutron powder diffraction on CD$_3$ND$_3$PbI$_3$; $^{d \text{ and } e}$Ref. \cite{Lepper_2019} calculated with a $GW_0$ including spin-orbit coupling approach using electronic properties determined at the PBE and PBE0 \cite{Adamo_1999} levels, respectively; $^f$Ref.\cite{Brivio_2014} determined with a $QSGW$ ($QS$ for quasi-particle self-consistent) approach including spin-orbit coupling using the LDA electronic properties, respectively; $^g$Ref. \cite{Mosconi_2016} calculated using the $G_0W_0$ approximation including spin-orbit coupling and using the PBE electronic properties; $^h$Ref. \cite{Mosconi_2014} calculated using the $G_0W_0$ approximation including spin-orbit coupling and using the PBE electronic properties; $^i$Ref. \cite{Ponce_2019_2,Ponce_2019} calculated using a self-consistent $GW$ approach including spin-orbit coupling from electronic properties at LDA level; $^j$Ref. \cite{Filip_2015} calculated using a self-consistent $GW$ approach including spin-orbit coupling from electronic properties at LDA level; $^k$Ref. \cite{Filip_2014} calculated using a self-consistent $GW$ approach including spin-orbit coupling from electronic properties at LDA level; $^l$Ref. \cite{Caputo_2019}; $^m$Ref.\cite{Maculan_2015}; $^h$Ref. \cite{Li_2016}; $^n$average value of experimental data from Table 2 and between parentheses range of variation of the band gap with different materials formings and 
measurement techniques cited in Table 1 of Ref. \cite{Das_2022}, respectively; $^o$Ref. \cite{Whitfield_2016} neutron powder diffraction on CD$_3$ND$_3$PbI$_3$; $^p$Ref. \cite{Lopez_2020} at RT; $^q$Ref. \cite{Lopez_20} at 140K.
\end{table*}

\noindent Let's focus on the cubic to tetragonal phase transition. Though the cell's deformation is very tiny (in the pseudocubic reference frame, $c/a$ is $\approx$ 0.99), the impact on $E_g$ and $\Delta H$ is not negligible (the band gap variation is about 0.25 eV for a $\Delta H$ of -200 meV); this is directly linked to the PbX$_6$ octahedra tilting as illustrated by the data of table \ref{tab:calc_crystal_MAPbI}. We determine the band gaps for the fully optimized tetragonal cell and for the tetragonal cell optimized keeping the iodine octahedra at their pristine position: the lattice parameters variations are less than 0.2\%, and, the MA cations ordering in both cells is quasi-equivalent; however the rotation of the iodine octahedra of $\approx$ 10$^\circ$ yield an increase of $E_g$ of 0.15 eV, and of $\Delta H$ of quasi 50\%.

\noindent These trends are more pronounced in the orthorhombic phases which allow both the tilting of halides octahedra and different orderings of the MA moities. The $\Delta H$ obtained between the cubic and the different orthorhombic phases are greater than those obtained between the cubic and tetragonal phases: The cell anisotropy in terms of lattice parameter variation with respect to the cubic one can reach 3\%; for any orthorhombic phase, the rotations of the halides octahedra are equivalent and can go to up to $\approx$ 15$^\circ$ with respect to the pristine perovskite; the differences between each of them comes from the different ordering of the MA moities inside the cells, for the considered orthorhombic phases in this work the one which minimizes $\Delta H$ is the one which minimizes the electrostatic dipole-dipole energy, here the $P222_1$ crystal structure.

\noindent As  discussed in \cite{JLi_2016,JLi_2018}, the combined effects of the octahedra tilting and MA ordering lead to the band gaps increase (this $E_g$ increase with the symmetries lowering is obtained whatever the level of approximations of the first-principles calculations, see table \ref{tab:calc_crystal_MAPbI}; the obtained increases of $E_g$ at the many-body perturbation theory level are lower than ours because the used cells are fixed at the experimental data and their size do not allow a full relaxation of halides octahedra and MA moities). As expected, it can be noticed that the MA orders in orthorhombic cells have a lowest impact on $E_g$ than the octahedra tilting: the difference of band gaps between each orthorhombic phases is less than 5\%.

\noindent The combination of the PBEx hybrid functional with our "nano-crystallographic" model of $\rm MAPbI_3$ allow to obtain a satisfactory agreement between the calculated geometries and band gaps, and, the literature; we now proceed to the study of the $\rm MAPbI_3$ phase impacts on its optoelectronics properties. 
\subsection{Optical indices}
\noindent Table \ref{tab:dielhamiltonian} provides the results obtained for further Hamiltonians on the band gaps, valence band edge energy and refractive index for the cubic phase of $\rm MAPbI_3$. They are compared to available experimental data. Once again, this illustrates the trends already discussed in the previous section: As expected PBE underestimates $E_g$ and $E_{VBM}$ by 40 and 10 \% with respect to the experimental data, respectively, while the hybrid functionals PBEx and HSE correct this discrepancy, the best agreement obtained for PBEx. For $\chi$ and $n$, the considered functionals provide data close to experiment (or included in the experimental dispersion or precision).
\begin{table}[!h]
\begin{center}
\caption{\label{tab:dielhamiltonian}Direct optical band gap $E_g$ (in eV), the valence band edge energy $E_{VBM}$ (in eV), electronic affinity $\chi$ (in eV), and refractive indices $n$ determined from electronic contribution of the dielectric tensor component $\varepsilon_{\infty}$ for the cubic $\rm MAPbI_3$ as obtained with the
PBEx hybrid functional (see text), PBE exchange-correlation functional \cite{Perdew_2008} and the HSE screened hybrid functional \cite{Krukau2006}. When available experimental data are given for comparison.}
\begin{tabular}{lcccc}
\hline
Method&$E_g$&$E_{VBM}$&$\chi$&$n$\\
\hline
PBEx&1.68&5.47&3.79&2.18\\
PBE&0.92&4.98&4.06&2.35\\
HSE&1.36&5.33&3.97&2.20\\
&&&&\\
Exp.&1.60$^a$, 1.70$^b$&5.05$^a$&3.45$^a$&2.29$^f$\\
&1.62 (1.50--1.69)$^c$&5.50$^d$&3.90$^d$&\\
&&5.92$^e$&4.10$^e$&\\
\hline
\end{tabular}
\end{center}
\footnotesize
$^a$Ref. \cite{Caputo_2019}; $^b$Ref. \cite{Li_2016}; $^c$average value of experimental data from Table 2 and between parentheses range of variation of the band gap with different materials formings and  measurement techniques cited in Table 1 of Ref. \cite{Das_2022}, respectively; $^d$Ref. \cite{Haque_2019} from figure 12; $^e$Ref. \cite{Olthof_2_2016} from figure 5; $^f$Ref. \cite{Yuan_2015} from figure 5.
\end{table}

\begin{figure}[h!]
\centering
a) \includegraphics[scale=0.2875]{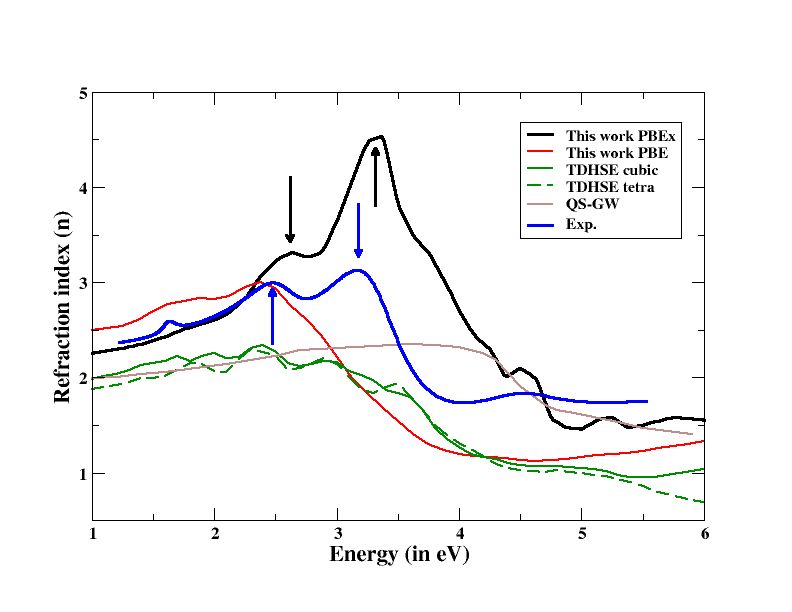}\\
b) \includegraphics[scale=0.2875]{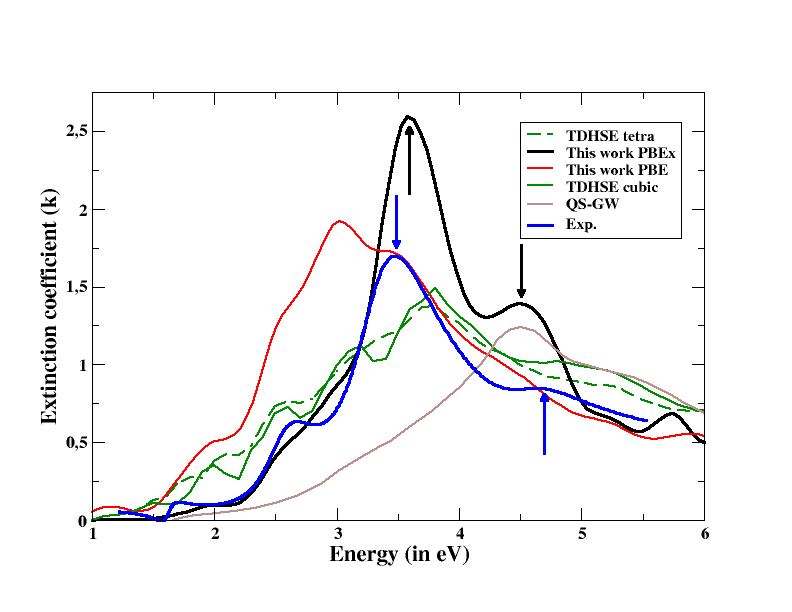}\\
\caption{Optical indices a) $n$ and b) $k$ determined at different level of approximations for the cubic phase of $\rm MAPbI_3$. In green, $n$ and $k$ deduced from the dielectric responses of the cubic and tetragonal cell description of Demchenko {\em et al.} determined at the TDHSE level \cite{Demchenko_2016}; in brown, $n$ and $k$ determined in the QS-$GW$ approximation on a single cubic cell \cite{Leguy_2016}. Experimental data from ellipsometry on single crystal \cite{Leguy_2016} are given for comparison.\label{fig:complexenk}}
\end{figure}

\begin{figure}[h!]
\centering
a) \includegraphics[scale=0.2875]{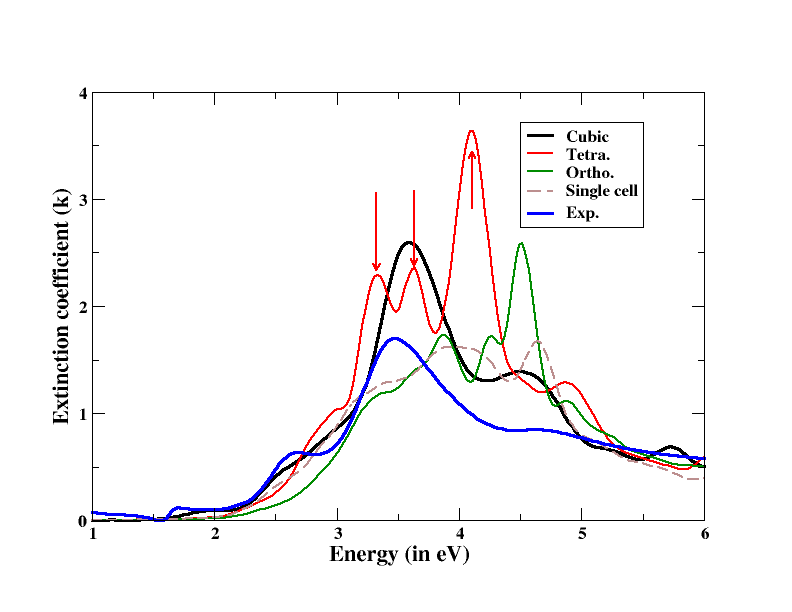}\\
b) \includegraphics[scale=0.2875]{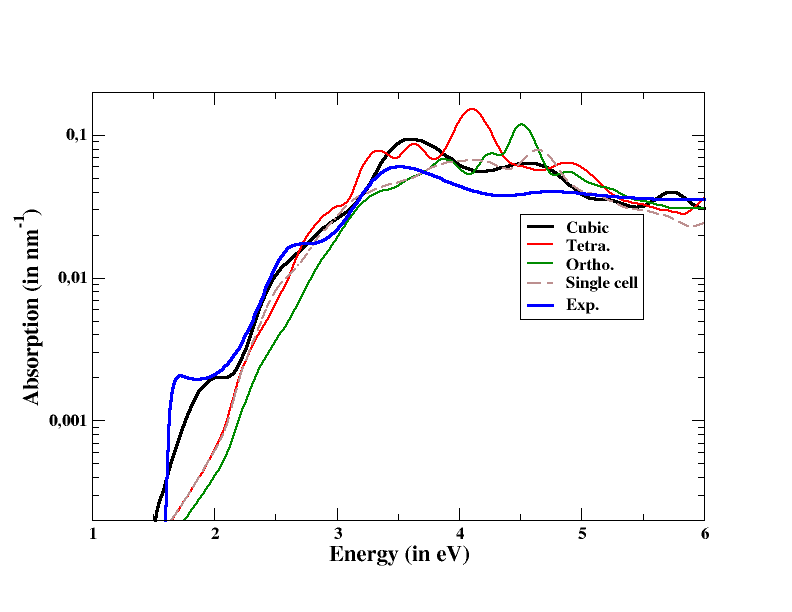}\\
\caption{Extinction coefficient a) and absorption b) determined for different phases of $\rm MAPbI_3$ at the PBEx level. The symmetries of the cubic, tetragonal, orthorhombic phases and single cell are $Pm\bar{3}m$, $I4/mcm$, $P222_1$ and $R3m$, respectively. Experimental data from ellipsometry on single crystal \cite{Leguy_2016} are given for comparison.\label{fig:absorption}}
\end{figure}

Let's focus on the refractive indices $n$ and extinction coefficients $k$ spectrum of the cubic $\rm MAPbI_3$. Figures \ref{fig:complexenk} and \ref{fig:absorption} represent the obtained results for different Hamiltonians on its spectral responses, and, the impacts of the phase changes (including single cell calculations) on these data, respectively. Following the same philosophy as in the previous sections, they are compared with theoretical data obtained at the QS-$GW$ \cite{Leguy_2016} and TDHSE \cite{Demchenko_2016} levels. Both of them use a "local" description of the system: at the QS-$GW$ level, $n$ and $k$ are determined by averaging the results obtained with the MA cation oriented along the $<0 0 1>$, $<1 1 0>$, and $<1 1 1>$ directions of a single cubic cell; the TDHSE calculations have been realized with a single cubic cell and a tetragonal cell described (a $\sqrt{2}$ $\times$ $\sqrt{2}$ $\times$ 2 supercell of the cubic cell) where all the atomic positions have been optimized. The experimental data choosen for comparison are obtained from ellipsometry measurements on single crystals \cite{Leguy_2016} to be coherent with our calculations.
 
\noindent Figures \ref{fig:complexenk} and \ref{fig:absorption}a) show that the various calculations based on "local" descriptions, even when performed using "state of the art" methods, fails to obtain an optical response that is qualitatively in good agreement with the experiment.

\noindent However, our crystallographic model combined with our tuned PBEx hybrid functional allows to significantly correct the obtained spectra: for $n$ and $k$, the shape of the spectrum is well reproduced as well as the main peak positions which are in good agreement with experiments. However the width and maximum of the main peaks are overestimated compared to the experimental spectra. These discrepancies might be related to further factors: 
\begin{itemize}
\item[i)] The lack of spin-orbit coupling which is not taking into account in our calculations, and/or, our functional which does not contain mid and long range interactions \cite{Demchenko_2016}. The current version of the CRYSTAL code do not permit the evaluation of the performance of the most sophisticated functional and spin-orbit coupling influences on the dielectric responses of materials.
\item[ii)] In a sample, statistical averaging of the MA distribution and their consequences on the crystal deformations and phase transitions could lead to smoother optical indices and absorption compared to the one calculated from an ordered static single crystal.
\end{itemize}
Nevertheless, our approach delivers results that can serve as a reference for evaluating the influence of phase changes on optical indices.
\indent Let's focus on the impact of the phase changes on $n$ and $k$ of $\rm MAPbI_3$. The table \ref{tab:dielphase} illustrates the induced anisotropy on the optical properties by these changes. It clearly shows that the halide octahedra distortions of phases lead to an increase of the band gaps and a decrease of the electronic affinity due to the combined effects of the down shift of the valence band and the up shift of the conduction band (see Fig. SI, the example of the tetragonal phases in table \ref{tab:dielphase} shows that keeping the PbX$_6$ octahedra undistorted leads electronic and optoelectronic properties near to the cubic ones as for the $P4/mmm$ tetragonal phase). This increases the anisotropy of the refractive indices, the $n$ variations in each direction of the crystals lattice thus reaching up to 8\%. Consequently, the tetragonal and orthorhombic phases are birefringent.
\begin{table*}[!h]
\begin{center}
\caption{\label{tab:dielphase}Calculated direct optical band gap $E_g$ (in eV), energy level of the valence band $E_{VBM}$ (in eV), electronic affinity $\chi$ (in eV), refractive indices tensor components  $n^{a, b, c}$ and  isotropic refractive index $\overline{n}$ (determined from electronic contribution of the dielectric tensor components $\varepsilon_{\infty}^{a, b, c}$) for different phases of $\rm MAPbI_3$ with the PBEx hybrid functional. For the tetragonal $I4/mcm$ phase, between parentheses are the obtained data for $\rm MAPbI_3$ optimized keeping the PbI$_6$ octahedra undistorted.}
\begin{tabular}{lccccccc}
\hline
Phase&$E_g$&$E_{VBM}$&$\chi$&$n^a$&$n^b$&$n^c$&$\overline{n}$\\
\hline
Cubic&&&&&&&\\
\scriptsize $Pm\bar{3}m$&1.68&5.47&3.79&2.18&&&2.18\\
&&&&&&&\\
\multicolumn{8}{l}{Tetragonal}\\
\scriptsize $P4/mmm$&1.70&5.41&3.71&2.17&&2.22&2.19\\
\scriptsize $I4/mcm$&2.17&5.65&3.49&2.12&&2.24&2.16\\
&({\em 1.81})&({\em 5.47})&({\em 3.66})&({\em 2.17})&&({\em 2.20})&({\em 2.18})\\
&&&&&&&\\
\multicolumn{8}{l}{Orthorhombic}\\
\scriptsize $Pnma$&2.35&5.73&3.38&2.08&2.14&2.14&2.12\\
\scriptsize $Pna2_1$&2.24&5.68&3.44&2.08&2.14&2.17&2.12\\
\scriptsize $P222_1$&2.25&5.70&3.46&2.07&2.11&2.17&2.12\\
&&&&&&&\\
\multicolumn{8}{l}{Single cell}\\
\scriptsize $R3m$&2.25&5.70&3.45&2.13&&2.18&2.15\\
\hline
\end{tabular}
\end{center}
\end{table*}
\\

Figure \ref{fig:absorption} represents the obtained results on the impacts of the phase changes on the extinction coefficient $k$ and the absorption spectra of $\rm MAPbI_3$.

\noindent Considering figure \ref{fig:absorption}a) and focusing on the differences between the cubic and tetragonal obtained spectra, it clearly shows: i) the splitting in two and the decrease of intensity of the peak at 3.5 eV, and, ii) the appearance of a third peak at 4. eV. The splitting and the decrease of intensity are linked to the lift of degeneracy between the $a$ and $c$ directions and the associated Pb-off center displacement inside the halide octahedra; the third peak is related to the tilting of the iodine octahedra.\\
For the orthorhombic phase, the situation is more complicated since the Pb-off center displacements, and, the tilting of the I-octahedra are allowed in all  lattice directions. The additional peak above 4ev always exists but with a smaller intensity. The blue shift of the  different peaks is associated to the increase of the band gap from the cubic, to the tetragonal and to the orthorhombic phases. The high intensity and discreet character of these additional peaks may be due to the lack of spin-orbit coupling and to our static ordered crystal model.

\noindent Nevertheless, this crystallographic model allows a detailed identification of the obtained spectra for a given phase which is not the case with a "local" description of the perovskite (for instance, the obtained optical properties by Demchenko {\em et al.} \cite{Demchenko_2016} for the cubic and tetragonal phases are similar, see Fig. \ref{fig:complexenk}).

Once again, absorption results in figure \ref{fig:absorption}b) show that better agreement with the experimental data is obtained for the cubic phases. The spectra of the other phases are characterized by the supplementary peaks above 3 eV. This does not mean that the local order in the sample is cubic since the measurement can be considered as an average description of the optical responses of the perovskite with a potential smoother absorption spectrum and the  differences observed between experiment and theory are satisfactory. Let's see the impact of these data on solar cell performance.
\subsection{Device performance}
In this section we investigate the device performance with the different theoretical and experimental data. The device performances are compared for the same structure in all cases, changing only the band parameters and optical functions detailed in the device modeling section. The main results on the short circuit current ($\rm J_{SC}$),  open circuit voltage ($\rm V_{OC}$), fill factor (FF), and efficiency ($\eta$) of the cell are summarized in the table \ref{tab:performance}.
\begin{table}[h!]
\caption{\label{tab:performance}Performance of devices for different phases of $\rm MAPbI_3$ and different levels of approximation.}
\begin{center}
\begin{tabular}{cccccc}
\hline
Phase&$E_g$&$\rm J_{SC}$    &$\rm V_{OC}$       &FF&$\eta$\\
&(eV)&$\rm(A/m^2)$    &$\rm(V)$       &(\%) &(\%)\\
\hline
\multicolumn{6}{l}{\bf PBEx level}\\
Cubic&1.68& 18.20&1.30&85.60 &20.3 \\
\scriptsize $Pm\bar{3}m$&&&&&\\
Tetra.&2.17&13.30&1.80&78.50&18.7 \\
\scriptsize $I4/mcm$&&&&&\\
Ortho.&2.25&11.70& 1.97& 77.45 & 17.0 \\
\scriptsize $P222_1$&&&&&\\
&&&&&\\
\multicolumn{6}{l}{\bf PBE level}\\
Cubic &0.92&35.60& 0.51&73.30&13.2 \\
\scriptsize $Pm\bar{3}m$&&&&&\\
&&&&&\\
\multicolumn{6}{l}{\bf TDHSE level$^a$}\\
Cubic&1.57& 28.10  &1.21  &85.10   & 28.8 \\
Tetra.&1.54& 27.60 & 1.17 &84.80   & 27.4 \\
&&&&&\\
\bf Exp.$^b$&1.55&17.40&1.17&91.00&18.6 \\
\hline
\end{tabular}
\end{center}
\footnotesize
$^a$The TDSHE results are deduced from the dielectric responses of the cubic and tetragonal cell description of Demchenko {\em et al.} \cite{Demchenko_2016}; $^b$Experimental performances are determined from the band gap and the ellipsometry measurments on single crystal of Leguy {\em et al.} \cite{Leguy_2016}.
\end{table}
\subsubsection{Experimental baseline}
\begin{figure}[h!]
\centering
a) \includegraphics[scale=0.2875]{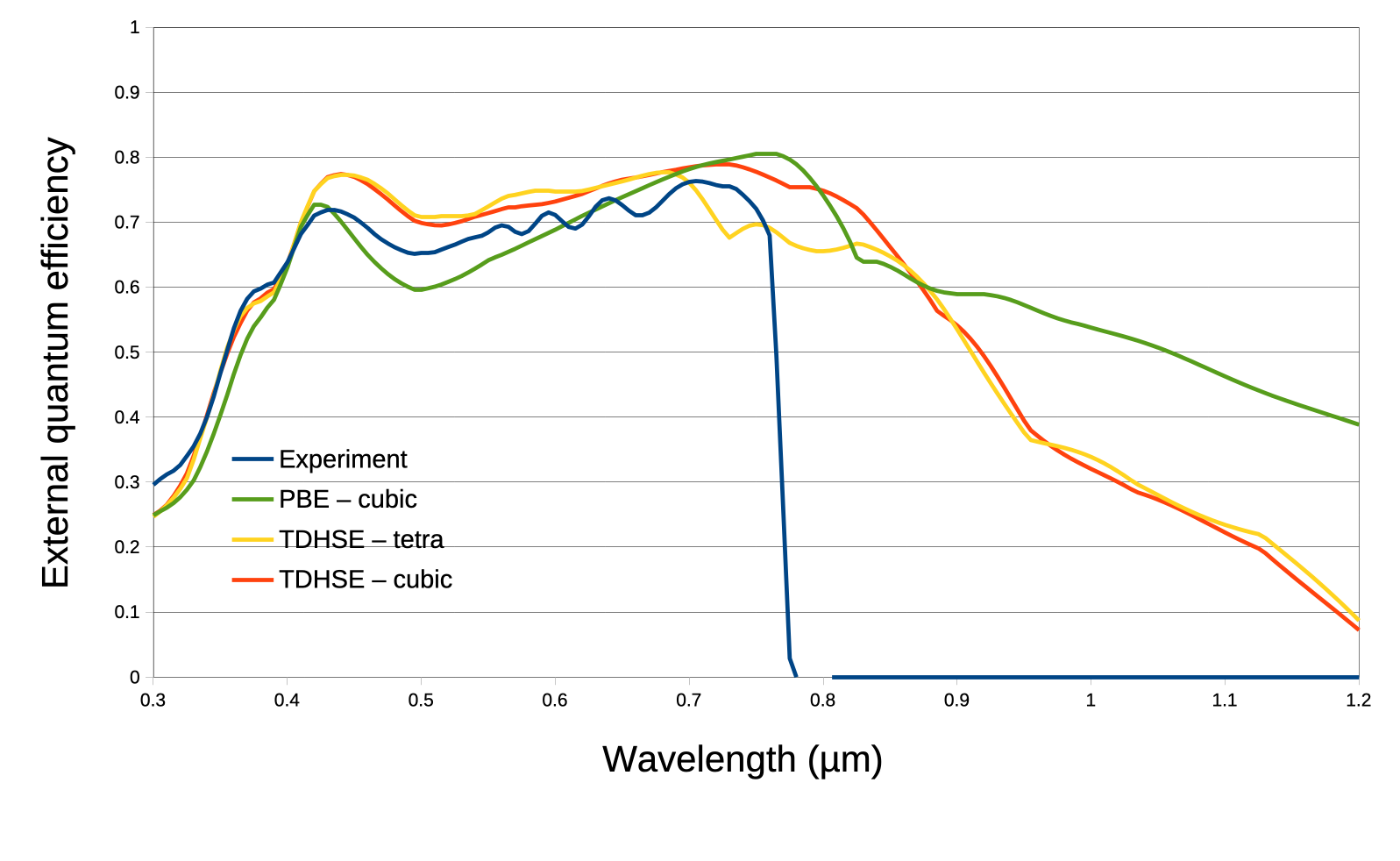}\\
b) \includegraphics[scale=0.2875]{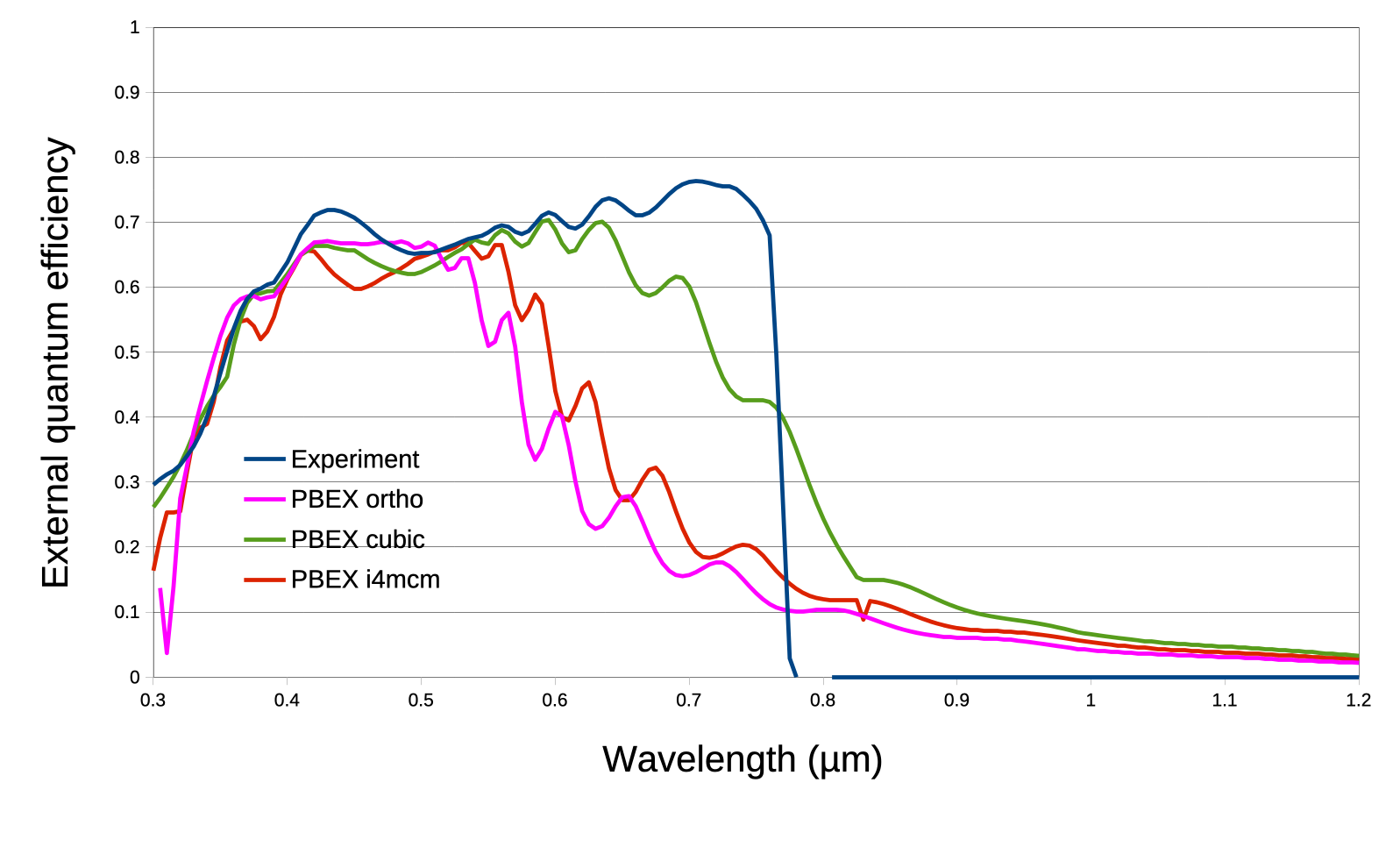}\\
\caption{Quantum efficiency obtained at the a) PBE and TDHSE and the b) PBEx levels. Quantum efficiency based on the same reference experimental data is given for comparison.\label{fig:device-model-EQE}}
\end{figure}

\noindent The experimental performance corresponds to a simulation using experimental optical and band parameters, just the same as the multiscale modeling which uses theoretical values of the same parameters. The corresponding simulated quantum efficiency is given in figure \ref{fig:device-model-EQE}. This shows a sharp cutoff which is a consequence of optical function smoothing of experimental data to remove below gap experimental artefacts. The calculated device current voltage characteristics under standard test conditions (STC)  are not shown for brevity, but determine the experimental parameters listed in table \ref{tab:performance}. We note in passing that the very high fill factor is just about physically possible in the radiative limit but it is beyond the scope of this paper to investigate this further.

 The efficiency of 18.6\% for this structure is however not high, which is a result of the lack of optimisation of layer dimensions and lack of an anti-reflection coating  in order to keep a comparable baseline for the different materials used. It would otherwise be necessary to optimise the device separately for each material model, which is not relevant for this study comparing theoretical modeling approaches.
\begin{figure}[h!]
\centering
a) \includegraphics[scale=0.2875]{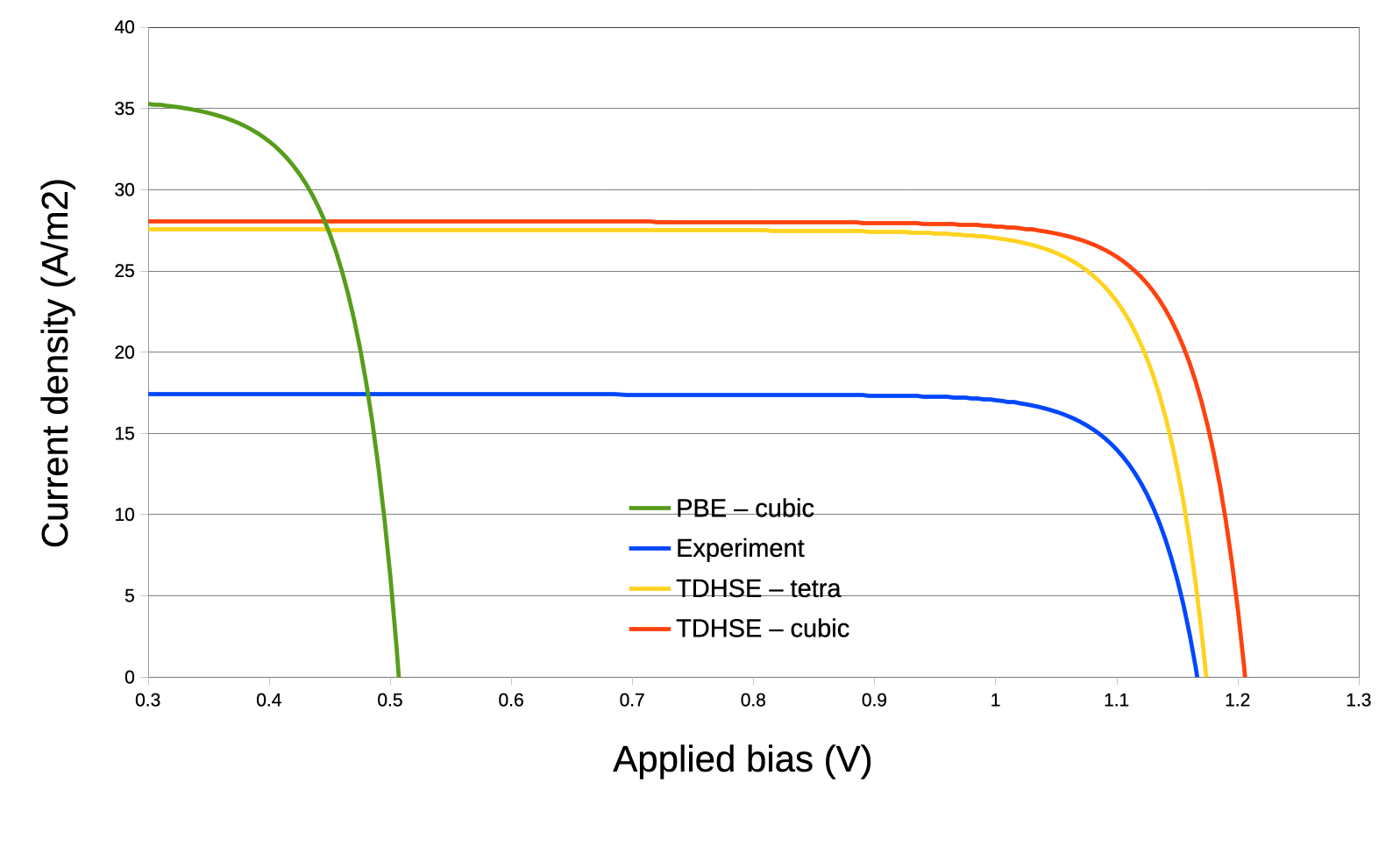}\\
b) \includegraphics[scale=0.2875]{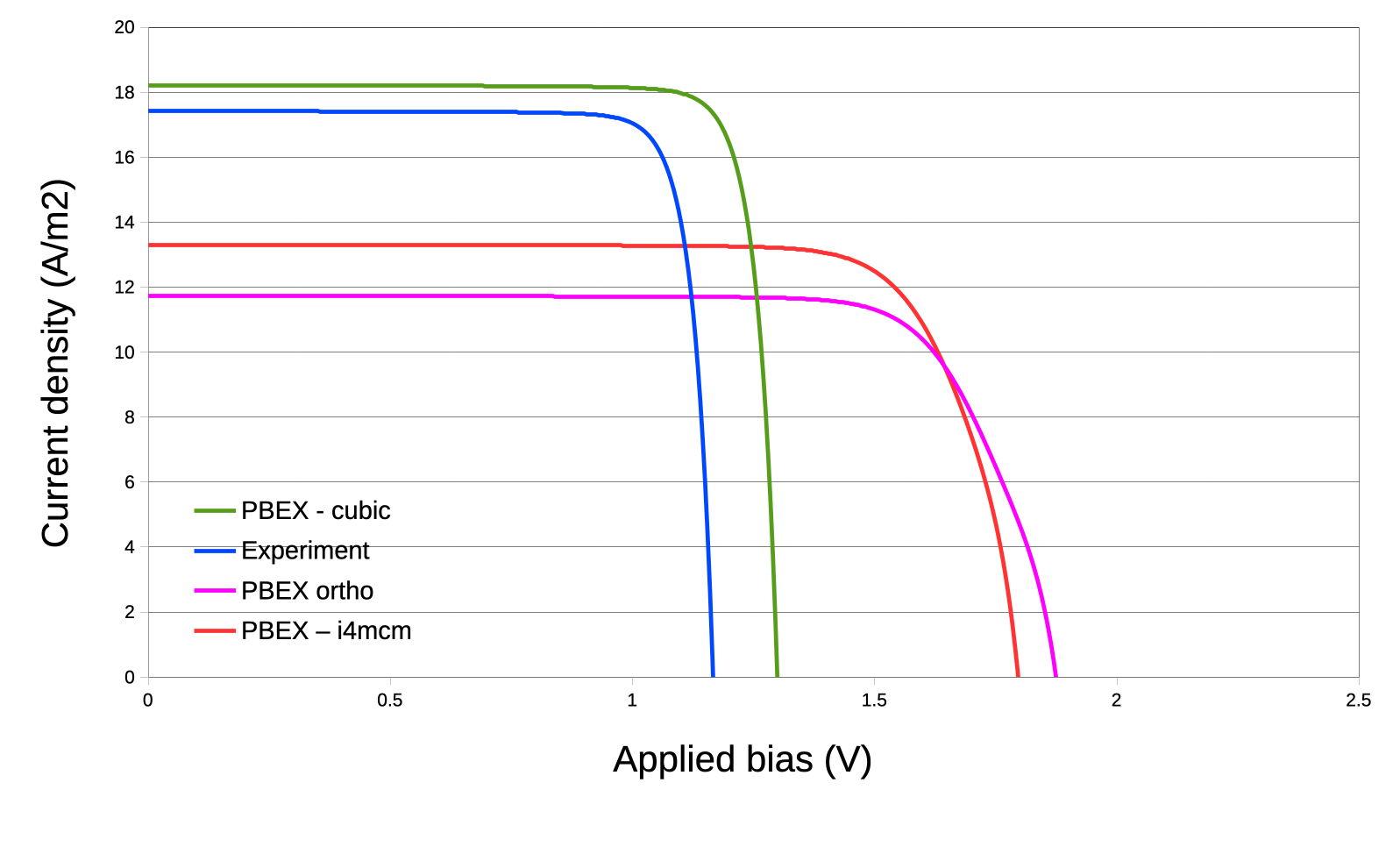}\\
\caption{Light current curves under AM1.5g for a) PBE and TDHSE, and, b) PBEx levels. Experimental current curve based on the same reference experimental data is given for comparison.\label{fig:device-model-LIV}}
\end{figure}
\subsubsection{Performance at the PBE and TDHSE levels}
\noindent We next look at the results obtained with the PBE functional for the cubic $\rm MAPbI_3$, and, at the TDHSE level with the optical responses of Demchenko {\em et al.} \cite{Demchenko_2016} for a "local" description of the cubic and tetragonal phases.\\
At the TDHSE level, the extinction coefficient is somewhat under-estimated (figure \ref{fig:complexenk}) in both cubic and tetragonal cases to slightly different extents. These two phases yield similar performance with slight difference in optical indices demonstrated by the quantum efficiency modeling in figure \ref{fig:device-model-EQE}(a). The modeled efficiency of these materials (table \ref{tab:performance}) is however very high at 27.4 \% and 28.8 \%, given that the radiative limiting efficiency at a band gap of 1.54 eV is 29.5 \%. This is clearly not physically possible given the quantum efficiency (QE) which is far lower than the unit QE which is one requirement in achieving the radiative limit, the other being that only radiative recombination loss mechanisms exist, which is not the case in this modeling which includes non-radiative recombination as we have seen in the device model section.

 Instead, we note in figure \ref{fig:device-model-EQE}a) that the absorption extends well below the band gap of these materials. This significantly over-estimates device performance since this device is delivering a current at a operating potential proportional to the gap, while absorbing photons with significantly lower energy which contradicts energy conservation, and yields excessively high efficiencies.

\begin{figure}[h!]
\centering
\includegraphics[scale=0.2875]{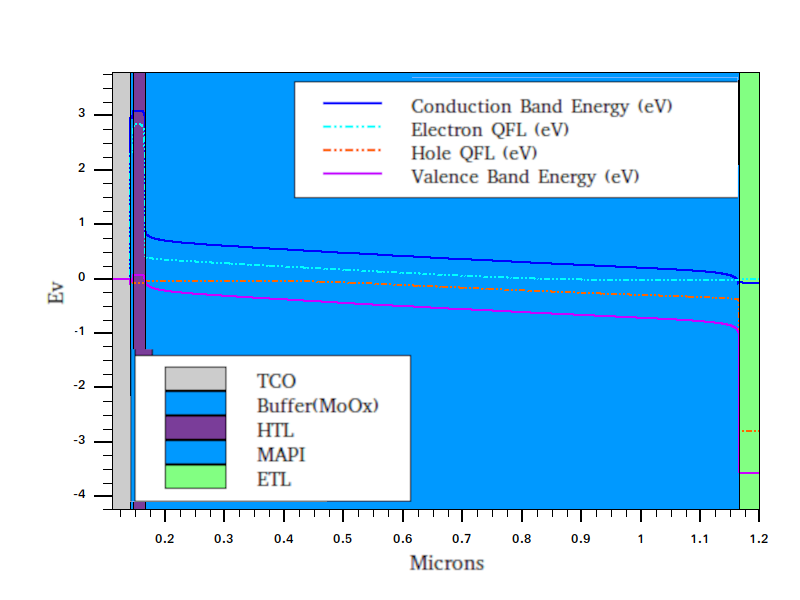}\\
\caption{Band profile and quasi-Fermi levels at the operating voltage for the cubic material at the PBE level showing a position dependent quasi-Fermi level separation and a low field consistent with a low fill factor and low efficiency. \label{fig:device-model-pbe-bands}}
\end{figure}

 Considering next the cubic PBE functional, with a band gap of 0.92 eV as shown in table \ref{tab:performance} which is significantly below the experimental value. The band profile of this device shown in figure \ref{fig:device-model-pbe-bands} shows however the the quasi-Fermi level separation in the space charge region varies with position. This means the space charge-region is departing from the case of constant quasi-Fermi level separation in good material, which is a consequence of a low field in this region. This results in a lower fill factor as seen in the light current curve  figure \ref{fig:device-model-LIV} and shown in table \ref{tab:performance}.
From these studies, we conclude that the PBE approximation have strengths in being able to phenomenologically describe different materials without requiring case by case tuning, which contributes to their popularity. But this strength has a cost in accuracy. This investigation of implementing PBE results at device scales shows the uncertainties in both electronic band parameters and optical parameters leads to large errors in device performance. These errors in some cases lead to unphysical results which can either underestimate or overestimate physical processes and device performance. We conclude that these methods as applied here are not suitable for the evaluation of device performance.
\subsubsection{Performance at the PBEx level}
\noindent We now look at three examples of the PBEx hybrid exchange-correlation functional method for three different phases. We see first that the quantum efficiency is under-estimated near the band gap, and shows fringes which are artefacts of the atomistic modelling and not optical in origin. We also see a long tail below the gap which will tend to increase the efficiency, in particular because the band gap remains close to the experimental value. Nevertheless, the overall error introduced by this low absorption above the gap and non-zero absorption below it do not significantly impact the performance of the cell.

 The parameters for these materials in table \ref{tab:performance} show open circuit voltages consistent with band gaps and an efficiency variation consistent with expected behaviour, with the lowest band gap having the highest efficiency since it is closest to the single junction optimum of 1.35 eV in the Shockley-Queisser limit. The simulated light current characteristics are shown in figure \ref{fig:device-model-LIV}(b). We again see a consistent behaviour with the lowest band gap material showing the highest $\rm J_{SC}$ and lowest $\rm V_{OC}$, which decrease and increase respectively for the two higher PBEx band gaps.

 This demonstrates a first method of evaluating materials parameters for different phases of a given material, in this case $\rm MAPbI_3$. The optical parameters and electronic band structure provided by the atomistic model to the device model yield results which are physically consistent. Aspects which require further attention are below band gap absorption tails, the absorption underestimation above the gap. The method is however sufficient to evaluate the potential performance of promising new perovskite photovoltaic materials. We conclude that this atomistic approach is, for the chosen material and its phases, is suitable for integration with device modelling for the chosen materials and phases, and can help guide the development of such materials for use in solar cells.
\section{Conclusions}
\noindent In conclusion, we present the first step of a pragmatic multiscale approach using atomistic scale first-principles calculations coupled to device scale numerical models. We have studied the impact of different methods of evaluating the optical properties of $\rm MAPbI_3$ properties at the first-principles level. A simple hybrid exchange-correlation functional combined with a crystallographic description enabling consideration of the MA ordering at the nano-scale have been optimized to yield description of its structural, electronic properties in good agreement with experiment. The obtained optical responses are in better qualitative agreement with experiment than the previously published ones. The  band gaps, affinities and optical indices obtained serve as input data to a device model to estimate the performance of solar cells.

 The preliminary device model results are also in qualitative agreement with experimental data. However, this methodology has to be demonstrated on a more detailed set of perovskites: further points have to be developed such as the effects of the spin-orbit coupling, the improvement of the hybrid functional, and, the adaptation of the atomic model to the experimental conditions (effect of the dimensionality (thin films) on the optical responses, for instance).

 If the trends are confirmed, it can  provide a set of criteria for optimizing the materials for different PV applications and for suggesting effective complex perovskites as well as other family of absorber.  In evaluating these impacts, we have the choice of two approaches concerning the device structure, that is, the properties of layers other than the absorber, such as the ETL and HTL properties for example. 
 
 The first approach is to fix all material parameters other than the absorber to those of the experimental structure. This provides a common baseline and has the benefit of clarity in that we look only at the effect of absorber optical and electronic properties. This is the approach followed in this paper. 
 
 The second approach is to optimise each structure independently in terms of absorber properties given by atomistic modeling. This approach has the benefit of providing a proper evaluation of the absorber material potential performance, but at the extent of losing the focus on comparison of absorbers, since the properties of the other layers in the device will vary. This will nevertheless be the focus of future work.
\funding
The authors thank the ANRT (French National Association for Research and Technology) for its financial support within CIFRE agreement 2023/0728 (industrial convention for training through research), and support from the France 2030 programme PEPR-TASE (“Programme et Equipements Prioritaires de Recherche sur les Technologies Avanc\'ees des Syst\`emes Energ\'etiques”) specifically within the MINOTAURE project, Grant ANR-22-PETA-0015.
\conflict
The authors declare no conflict of interest.
\dataavailability
Crystallographic description of the $\rm MAPbI_3$ different phases considered in this work, optimized structures and band gaps for different phases of MAPbX$_3$ (X = Cl, Br and I).
%
%
%
%
%
%
%
%
%
%
%
%
%
%
\bibliographystyle{edpsci}
\bibliography{edpsci}
\end{document}